\documentclass[useAMS,usenatbib,twocolumn, fleqn]{mn2e} \usepackage{natbib,
graphics, epsfig, color}

\usepackage{times}
\usepackage{amsmath}
\usepackage{amssymb}
\usepackage{textcomp}
\usepackage{verbatim}

\title[substructure profiles in lensing]{On the density profile of dark matter substructure in gravitational lens galaxies}
\author[S. Vegetti \& M. Vogelsberger]{Simona
  Vegetti$^{1}$ and Mark Vogelsberger$^{2}$\\
  $^1$Max Planck Institute for Astrophysics, Karl-Schwarzschild-Strasse 1, 85740 Garching, Germany\\
  $^2$Kavli Institute for Astrophysics and Space Research, Massachusetts Institute of Technology, Cambridge, MA 02139, USA\\}

\begin{document}

\maketitle

\begin{abstract}
We consider three extensions of the Navarro, Frenk and White (NFW) profile and investigate the
intrinsic degeneracies among the density profile parameters on the 
gravitational lensing effect of satellite galaxies on highly magnified Einstein rings. In particular, we
find that the gravitational imaging technique can be used to exclude specific regions of the considered parameter space, 
and therefore, models that predict a large number of satellites in those regions. 
By comparing the lensing degeneracy with the intrinsic density profile degeneracies, we show that theoretical predictions based on fits that are dominated
by the density profile at larger radii may significantly over- or under-estimate the number of satellites that are detectable with gravitational lensing.
Finally, using the previously reported detection of a satellite in the gravitational lens system JVAS\,B1938+666 as 
an example, we derive for this detected satellite values of $r_{\rm{max}}$ and $v_{\rm{max}}$ that are, for each considered profile, consistent within 
1$\sigma$ with the parameters found for the luminous dwarf satellites of the Milky Way and with a mass density slope $\gamma < 1.6$. 
We also find that the mass of the satellite within the Einstein radius as measured using gravitational lensing is stable against assumptions on the substructure profile.
In the future thanks to the increased angular resolution of very long baseline interferometry at radio wavelengths and of the E-ELT in the optical we will be able to set tighter constraints 
on the number of allowed substructure profiles.
\end{abstract}


\section{Introduction}
Observations of the Milky Way satellite galaxies have long been used as test
laboratories of the cold dark matter (CDM) paradigm in the small non-linear
regime. These have revealed a number of potential problems that include {\it the missing satellite problem} \citep[][]{Klypin1999,Moore1999}, {\it the core-cusp problem} \citep[][]{Moore1994, Kuzio2008, deBlok2010,
Walker2011, Amorisco2012} and {\it the too big to fail problem}
\citep[][]{Boylan2011,Boylan2012}. While some of these small-scale issues could potentially be
solved by baryonic feedback processes and star formation
\citep[][]{Governato2010,Governato2012,Brooks2012}, they may also be the
signature of a different physics of the dark matter \citep[e.g.][]{Lovel2012,
Vogelsberger2012}. 

As detecting satellite galaxies and measuring their
properties can be observationally challenging, most of the observations have
been limited to the Local Group, which may not necessarily be a fair
representation of the Universe. It is therefore important to extend these
observations to the satellite galaxies of other massive parent galaxies. \citet{Vegetti09a,Vegetti09b} have shown how highly magnified Einstein rings can be used
to detect faint satellites in gravitational lens galaxies out to any lens
redshift, and how the measured properties of these satellites can be then used to
constrain the mass function of mass substructure. This gravitational imaging technique detects mass
substructure in lens galaxies via their gravitational effect on the surface
brightness distribution of highly magnified Einstein rings and arcs. During this process, substructures are initially detected and their masses
measured in a substructure-model independent way, that is, via pixelated potential corrections to a smooth potential model for the parent halo; subsequently, the data are remodelled within
the context of a particular substructure model, and the most probable {\it a
posteriori} values of the model parameters (e.g. the substructure mass,
Einstein radius and position) given the data are determined. Generally, a
singular isothermal sphere (SIS) or a truncated isothermal sphere is used for
the substructure density profile. However, any other substructure model could in principle be
used provided it produces the same gravitational lensing effect and fits the data equally well.
Traditionally, substructures have been detected in lens galaxies via their effect on the magnification of multiply imaged quasars (i.e. flux-ratio anomaly).
However, due to the point-like nature of the lensed quasars, this type of data is not sensitive to individual substructures but to the general population.
Flux ratio anomalies, therefore, cannot be used to constrain the substructure mass function slope nor the density profile of each substructure, but can only be used 
to quantify the amount of substructures \citep{Dalal02} or constrain the concentration of their mass density profile \citep{Xu12} in a statistical sense.
In first approximation, gravitational lensing only provides a good measure of the mass within the
Einstein radius of the lens. However, in combination with other observables (e.g. stellar dynamics and weak lensing) or if the lensed images have extended radial structure, a value for the \emph{mean}
density slope between the radial extend of the observations can also be derived \citep[e.g.][]{Koopmans09,Barnabe09,Newman13,Grillo13}. 
At the scale of satellite galaxies \citet{Suyu10} have shown that if a satellite galaxy is located close to a lensed arc with a radial extent larger or comparable to the size of the galaxy, the morphological structure of the arc contains important information that allows us to constrain the mass distribution of the satellite. This is due to the fact that the satellites can affect the surface brightness distribution of the lensed images over their full radial extent.
At the same time, however, it has been shown by \citet{Schneider13} that due to the mass sheet degeneracy, combined with the lack of constraints over large 
regions of the lens plane, it is essentially not possible to measure the density profile of the deflector using only gravitational lensing.

In this Paper, we investigate the effect of different substructure density profiles on the surface brightness distribution of extended arcs. The aim is not to precisely measure the density profile of substructures but to quantify the degeneracies among different models and identify those regions of the profile parameter space that cannot reproduce the observed perturbation, hence set constraints on dark matter and/or galaxy formation models that predict a large number of subhaloes/satellites in those excluded regions. In practical terms, given that we know a SIS substructure profile to be a good description of the data (although not necessarily of the true underlying profile of the substructure), we will look for combinations of profile parameters that  provide an equally good fit and for those that are instead excluded by the data. With this aim we consider different extensions of the NFW profile. Since we are studying the lensing effect of small substructures, which are more likely dark matter dominated, we believe this choice of profiles to be well justified.
This method could provide an important test for models of dark matter. In Section \ref{sec:effect}, we define the lensing signature of satellites on gravitationally lensed images. In Section \ref{sec:mass_model}, we present the considered satellite mass models and their intrinsic degeneracies. Finally, in Sections \ref{sec:discussion} and \ref{sec:conclusions} we discuss our results and summarize our main conclusions, respectively.

Throughout we assume a flat cosmology with $\Omega_{\rm m}=$~0.25, $\Omega_\Lambda=$~0.75 and $H_0 =$~73~km\,s$^{-1}$~Mpc$^{-1}$. 

\begin{table}
\begin{tabular}{llllll}
\hline
Profile         &$\sigma_v$ [$\mathrm{km\,s}^{-1}$]     &\!\!\!\!$r_s$ [$\mathrm{kpc}$]   &$\gamma$                   &$r_c$ [$\mathrm{kpc}$]       &$r_t/r_s$\\
\hline
gNFW            &15.6                                   &\!\!\!\!0.1-5                    &0-2                        &-                            &-\\

gNFW$_2$        &10-50                                  &\!\!\!\!$R_{200}/c_{200}$        &0-2                        &-                           &-\\

gtNFW           &15.6                                   &\!\!\!\!0.1-5                    &0-2                        &-                            &1-3\\

gtNFW$_2$       &10-50                                  &\!\!\!\!$R_{200}/c_{200}$        &0-2                        &-                            &1-3\\

cNFW            &15.6                                   &\!\!\!\!0.1-5                    &1                          &0.1-5                        &-\\

cNFW$_2$        &10-50                                  &\!\!\!\!$R_{200}/c_{200}$        &1                          &0.1-5                        &-\\
\hline
\end{tabular}
\caption{The prior limits on the substructure
parameters (velocity dispersion $\sigma_v$, scaling radius $r_s$, inner density
slope $\gamma$, core radius $r_c$ and truncation radius $r_t$) for all of the mass
density profiles considered: a generalized NFW profile (gNFW), a generalized NFW
profile with a given mass-concentration relation (gNFW$_2$),  a generalized
truncated NFW profile (gtNFW), a generalized truncated NFW profile with a given
mass-concentration relation (gtNFW$_2$), a cored NFW profile (cNFW) and a cored NFW
profile with a given mass-concentration relation (cNFW$_2$).}
\label{tbl:mass_model}
\end{table}

\section{Lensing effect}
\label{sec:effect}
Given the surface brightness distribution $I(x)$ of the lensed images as a function of
the position $x$ on the lens plane, we define the effect of a given
substructure as the difference in $I(x)$ between a lens that contains the
substructure and the same smooth lens without the substructure, more precisely
we consider  
\begin{equation}
{\cal D}=  \sum_x{\left(\frac{B~I_{\rm smooth}(x)-(B~I_{\rm sub}(x)+n(x))}{2 ~\sigma(x)}\right)^2}\,.
\label{equ:effect}
\end{equation}
Here, $B$, is the blurring operator that encodes the effect of the telescope point-spread function
and $n(x)$ is the observational Gaussian noise with standard deviation $\sigma(x)$. 
In general, the sensitivity of the gravitational imaging technique depends on
the observational conditions as well as on the surface brightness distribution
of the background source, and the substructure mass and position relative to the
extended Einstein ring or arc.

Here, we test the specific case where all of these
variables are matched to those found from an analysis of Keck adaptive optics imaging of the gravitational lens system JVAS\,B1938+666 \citep{Laga12}, where a substructure of mass $\sim$10$^8$~M$_\odot$ was detected
with the gravitational imaging technique at the 12$\sigma$ confidence
level \citep{Vegetti12}. We leave a broader analysis of more general data properties, source
models and substructure masses and positions to a forthcoming paper.

\begin{figure}
\includegraphics[ width=8cm]{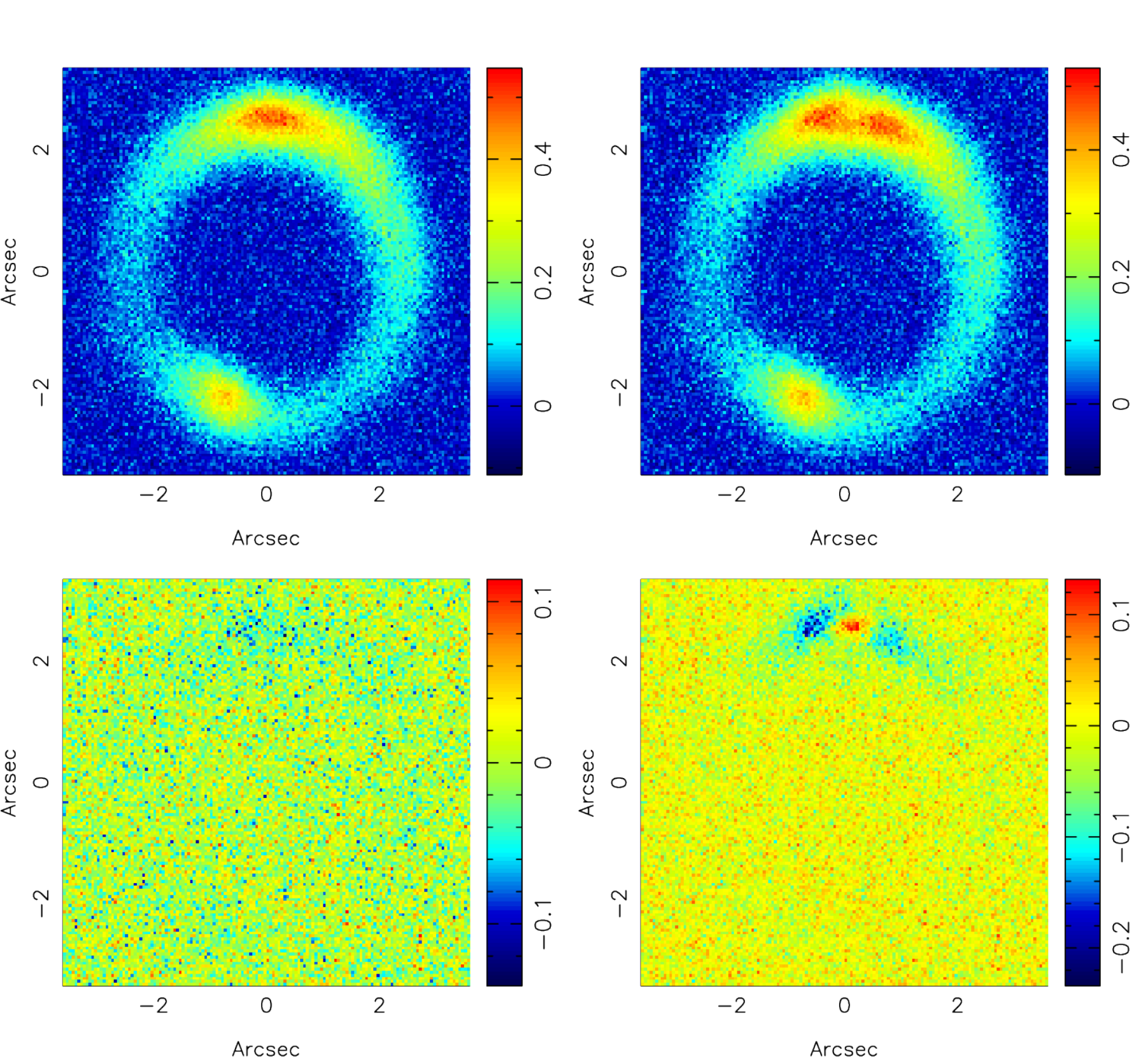}
\caption{Top panels: mock data of the gravitational lens system JVAS\, B1938+666 with a SIS substructure (left) and a gNFW substructure (right) with the same mass, slope $\gamma=0$ and scaling radius $r_s = 0.1$ kpc.
Bottom panels: difference in the lensed images surface brightness distribution between a smooth version of the gravitational lens system JVAS\, B1938+666 and a substructures version with a SIS substructure (left) and a gNFW substructure (right) with the same mass, slope $\gamma=0$ and scaling radius $r_s = 0.1$ kpc.}
\label{fig:profile_ex} 
\end{figure}

It should be noted that the lensing effect as defined in Equation (\ref{equ:effect})
is essentially a $\chi^2$,  and is therefore proportional to the log-likelihood of the data ($B~I_{\rm sub}(x)+n(x)$) given a smooth 
model ($B~I_{\rm smooth}(x)$). It is known that this quantity can
potentially lead to an over-estimate of the sensitivity to substructure of different masses and profiles.
As it is shown by \citet{Vegetti14}, the effect of a substructure can be partly re-absorbed with a change
in the source surface brightness structure and/or with a change in the lens macro model.
This implies that the most rigorous way to quantify the sensitivity of a given substructure (model)
is to compute the Bayes factor and marginalize over the lens and source parameters for 
a large number of mock data sets, each defined by a different combination of substructure mass models.  
Since this approach is computationally prohibitive and since in this Paper we are only interested in testing whether 
there is any observational signature to the mass density profile of the substructure under the simplest assumptions,
we limit our study to consider the $\chi^2$ above and refer to a future paper for a more rigorous quantification.
Indeed, if the $\chi^2$ is not sensitive to different mass models, this would be even more true for the Bayesian evidence.
As an example, we re-model the mock data shown in the top right panel of Fig. \ref{fig:profile_ex} with a SIS and a gNFW profile. By re-optimizing for the source, the main lens and the substructure parameters, we calculate the Bayesian evidence (${\cal{E}}$) of both models and find that the gNFW model is preferred by a $\Delta\log{\cal{E}}\sim$120; this corresponds roughly to a 15$\sigma$ significance. The likelihood ratio, related to equation (\ref{equ:effect}), is instead about 40 times larger (also in favour of the gNFW model). As expected, the degeneracy between the substructure, the lens macro model and the source structure can re-absorb the effect of a given substructure and reduce the difference among different profiles.
In practice, we expect the shape of the degeneracy among several profiles to be essentially unaffected, while the size of  the parameter region not compatible with the data to shrink or increase. Specifically, for a given substructure mass and position, the size of the profile parameter space that is excluded by the data, at a given level of significance, is set  by a combination of the data quality and the degeneracy between the substructure model and the macro model. Even though, in this Paper, we have only used an approximation for the latter, our test shows our current analysis to be robust for combination of profile parameters with lensing effects that are significantly larger or smaller than the effect of the reference SIS profile. Moreover, we expect the loss of sensitivity due to re-absorption by the macro model to be compensated, in the future, by a gain in sensitivity thanks to a higher angular resolution of the data (e.g. VLBI and E-ELT) and the sizes of the allowed and dis-allowed regions to be close to that derived in this more simple approach.

\section{Substructure mass model} 
\label{sec:mass_model}

\subsection{Parameterized mass models}

For gravitational lens modelling, the most widely used parametrization of the total mass density profile is the SIS, which has a three-dimensional density distribution $\rho\left(r\right)$ given by
\begin{equation}
\rho\left(r\right) = \frac{\sigma_v^2}{2 \pi G r^2}\,,
\label{eq:sis}
\end{equation}
where $\sigma_v$ is the velocity dispersion, $r$ is the radius and $G$ is the gravitational constant. As well as the SIS, we also consider substructure models with three other different mass density profiles that are variations of the \citet*[NFW;][]{NFW96} model. They are the generalized NFW (gNFW), the truncated generalized NFW (gtNFW) and the cored NFW (cNFW) profiles.

A generalized formulation of the NFW density profile (gNFW) of arbitrary inner slope
$\gamma$ was introduced by \citet{Zhao96},
\begin{equation}
\rho\left(r\right) =\frac{\rho_s}{\left(r/r_s\right)^\gamma\left(1+r/r_s\right)^{3-\gamma}}\,,
\label{eq:gnfw}
\end{equation}
where $r_s$ is the scaling radius at which the slope of the density profile
changes. The scaling radius is related to the concentration parameter
$c_{200}$, introduced by \citet{NFW96}, and to the radius $r_{200}$ that encloses a 
mass $M_{200}$, which has an over density of $200$ above the critical density $\rho_c$, and is defined by
\begin{equation}
r_s = \frac{r_{200}}{c_{200}} = \left(\frac{3\,M_{200}}{800\,\pi\,\rho_c\,c_{200}^3}\right)^{1/3}\,.
\end{equation}
For $\gamma=$~1, this reduces to the classical NFW profile, while for $\gamma=$~0
we have an inner cored profile and for $\gamma=2$ we have an isothermal profile.

\citet{Baltz09} introduced a smoothly truncated NFW profile, that we extend
here to the generalized case (gtNFW),
\begin{equation}
\rho\left(r\right) =\frac{\rho_s}{\left(r/r_s\right)^\gamma\left(1+r/r_s\right)^{3-\gamma}\left(1+(r/r_t)^2\right)^n}\,.
\label{eq:truncated}
\end{equation}
Here, $r_t$ is the truncation radius and $n$ sets the sharpness of the
truncation. In particular, we consider $n=$~1 and $r_t\geq r_s$.

We also consider a cored NFW profile (cNFW), as defined by \citet{Pena12} to be
\begin{equation}
\rho\left(r\right) =\frac{\rho_s}{(r/r_s+r_c/r_s)(1+r/r_s)^2}\,.
\label{eq:cored}
\end{equation}
In this case, $r_c$ is the core radius, and for $r_c=$~0 this profile reduces to the
classical NFW.

Generally, the main lensing properties, that is, the deflection angle $\alpha(x)$ and
the surface mass density $\Sigma(x)$ of these profiles do not always have
analytic expressions. These however, can be calculated numerically by solving for
the integrals of the surface mass density, 
\begin{equation}
\Sigma(x)= 2r_s\int_0^\infty{\!\!\!{\rm d}z~\rho\left(\sqrt{x^2+z^2}\right)},
\end{equation}
and the projected cylindrical mass,
\begin{equation}
M_{\rm{cyl}}(x) = 2\pi r_s^2\int_0^x{\!\!\!{\rm d}x^\prime~x^\prime~\Sigma\left(x^\prime\right)}\,.
\end{equation}
For a critical surface mass density,
\begin{equation}
\Sigma_c = \frac{c^2 D_s}{4\pi G D_{ds}D_d}\,,
\end{equation}
as a function of the angular diameter distances $D_s$ to the source, $D_d$ to
the lens and $D_{ds}$ between the lens and the source, the deflection angle is given by
\begin{equation}
\alpha(x) =\frac{ M_{\rm{cyl}}(x)}{\pi\,\Sigma_c\,x}\,.
\end{equation}
For all of the considered profiles, we set the normalization $\rho_s$ by
imposing a $M_{200}$ mass equal to that of a SIS of given velocity dispersion
$\sigma_v$,
\begin{equation}
M_{200}=\sigma_v^3\left(\frac{3}{800\,\pi\,\rho_c}\right)^{1/2}\!\!\!\left(\frac{2}{G}\right)^{3/2}\!\!\!\!\!=\int_0^{r_{200}}{\!\!\!\!\!\!\!{\rm d}r~4\pi r^2 \rho(r)}.
\label{eq:mass_vel}
\end{equation}
In this way, we ensure that all the substructures have the same mass and that
the comparison is meaningful.  For the truncated profile of Equation
(\ref{eq:truncated}), the normalization is obtained by integrating the
non-truncated version of the profile. This can be done because we expect the $M_{200}$ mass to be
set at a time prior to the accretion of the satellite by the host galaxy. Once
the normalization is fixed, all of the above models are left with two free
parameters, which are sometimes degenerate with each other, as for example
$r_s$ and $\gamma$ for the gNFW case. 

Some of the considered values of the scaling radius can result in substructures that are
more concentrated than the sub-haloes of the same mass that are seen in {\it N}-body CDM simulations. We
therefore also consider cases where the velocity dispersion (and hence mass) changes,
and the concentration is coupled to the mass by the 
mass-concentration relation defined by \citet{Duffy08},
\begin{equation}
c_{200}  =  5.71\left(\frac{M_{200}}{2 \times10^{12}~h^{-1}M_\odot}\right)^{-0.084}\left(1+z\right)^{-0.47}\,.
\end{equation}

The 12$\sigma$ significance level detection of a mass substructure in the 
gravitational lens galaxy JVAS\, B1938+666 at $z=0.881$ was obtained in a model independent way \citep{Vegetti12}. This detection was shown to be
consistent with a SIS that has $\sigma_v=$~15.6~km\,s$^{-1}$. In all of the considered cases given above, 
we set the substructure velocity dispersion and redshift to these measured values, and investigate which combination of
mass model parameters can result in a lensing effect (as defined by Equation \ref{equ:effect}) that is equal or comparable to 
what is produced by a SIS with these properties.

\subsection{Intrinsic degeneracies}
While Equation (\ref{equ:effect}) can be used to quantify the degeneracy between the lensing effect and different substructure models, 
each of the above profiles suffers from intrinsic degeneracies in their main parameters.
For example, the gNFW and the gtNFW can result in a high central density by having either a steep central slope $\gamma$ or a
large concentration parameter (i.e. small values of $r_s$), while the same effect can be achieved for a cNFW with either a small core or
a large concentration. Following \citet*{Wyithe01}, we quantify these intrinsic degeneracies by identifying those combinations of the $\rho(r)$ parameters
that minimize the following relations, relative to a set of reference profiles $\rho_0$ of given slope $\gamma$ (or core radius $r_c$) and
 different scaling radii $r_s$:
\begin{equation}
\chi^2_{rel} = \int_0^{r_{200}}{r^2\left(\frac{\rho(r)-\rho_0(r)}{\rho_0(r)}\right)^2{\rm d}r}\,
\end{equation}
and
\begin{equation}
\chi^2 = \int_0^{r_{200}}{r^2\left(\rho(r)-\rho_0(r)\right)^2{\rm d}r}\,.
\end{equation}
Specifically, for the gNFW and the gtNFW profiles, the reference profile
 $\rho_0$ is, respectively, a gNFW and a gtNFW with a slope fixed at $\gamma\equiv$~1 and with a scaling radius that is
 variable between the prior limits given in Table 1. For a cored NFW, $\rho_0$ is also a cored NFW with a fixed core 
 of $r_c =$~2~kpc and with a scaling radius that is variable within the same prior range. As already pointed out by \citet{Wyithe01}, the minimization of the $\chi^2$ is dominated by the central regions, so that a good fit is obtained
for the central density at the cost of a poorer fit at larger radii. Conversely, the minimization of the $\chi^2_{rel}$ is dominated by the density at larger distances
from the centre. For the truncated cases, the $\chi^2_{rel}$ is dominated by the regions around the truncation radius.
In the attempt of providing predictions or performing comparisons between theoretical expectations and gravitational 
lensing observations based on analytic fits to the mass density distribution of numerically simulated subhaloes, it is, therefore, very important to choose the proper fitting metric. Sub-optimal definitions of
the minimizing function (i.e. $\chi^2_{rel}$  versus $\chi^2$) can, in fact, lead to under-estimations or over-estimation of the subhalo gravitational lensing effect (see Section 4 and Fig. 2), and hence to biased results.
In particular, the correct comparison metric should be expressed in terms of deflection profiles [i.e. $M(<r)/r$] since this is the quantity that gravitational lensing is constraining to first order.

\begin{figure*}
\centering
\hspace{-0.1cm}\includegraphics[ width=0.2\textwidth]{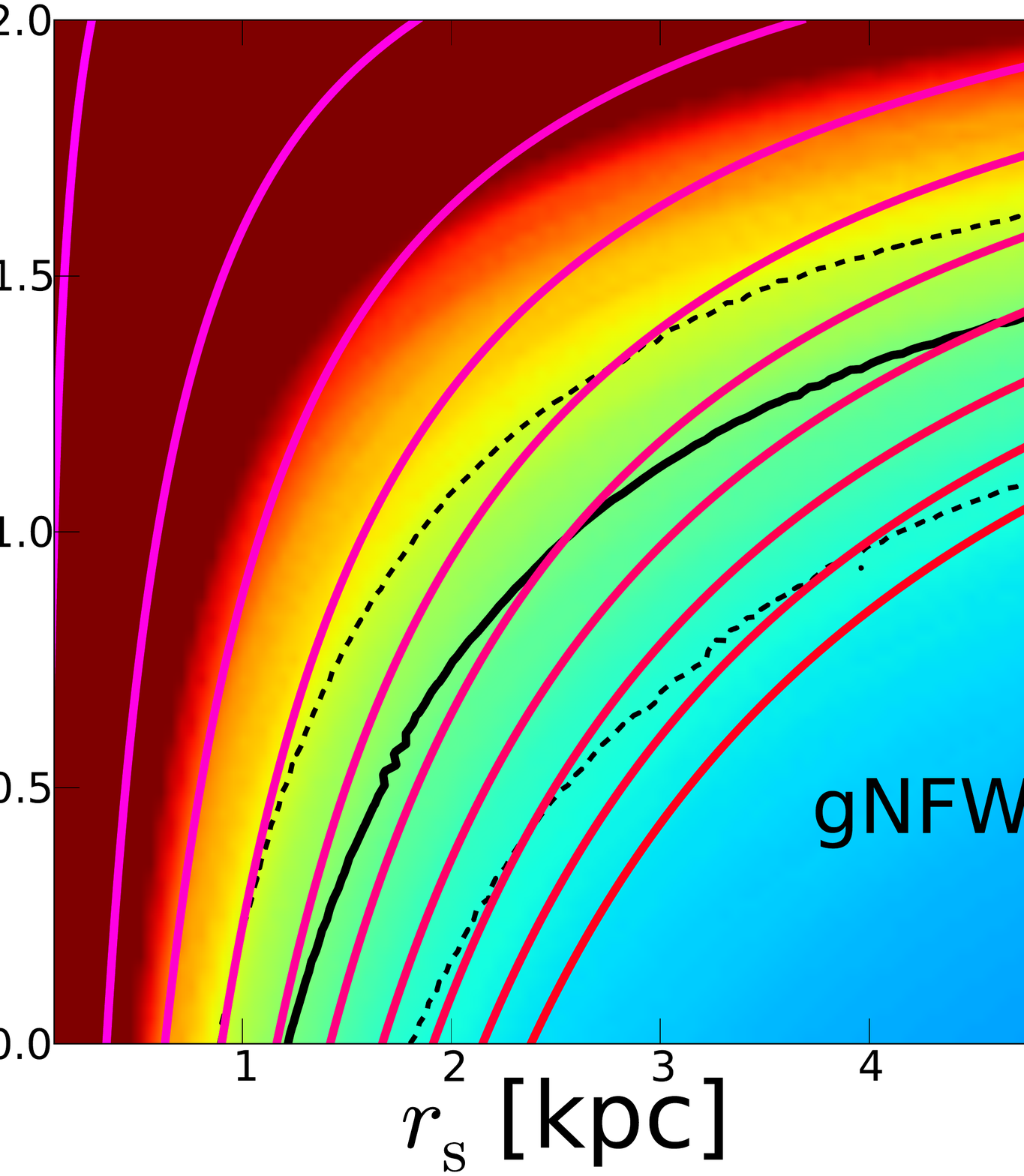}
\hspace{-0.1cm}\includegraphics[width=0.2\textwidth]{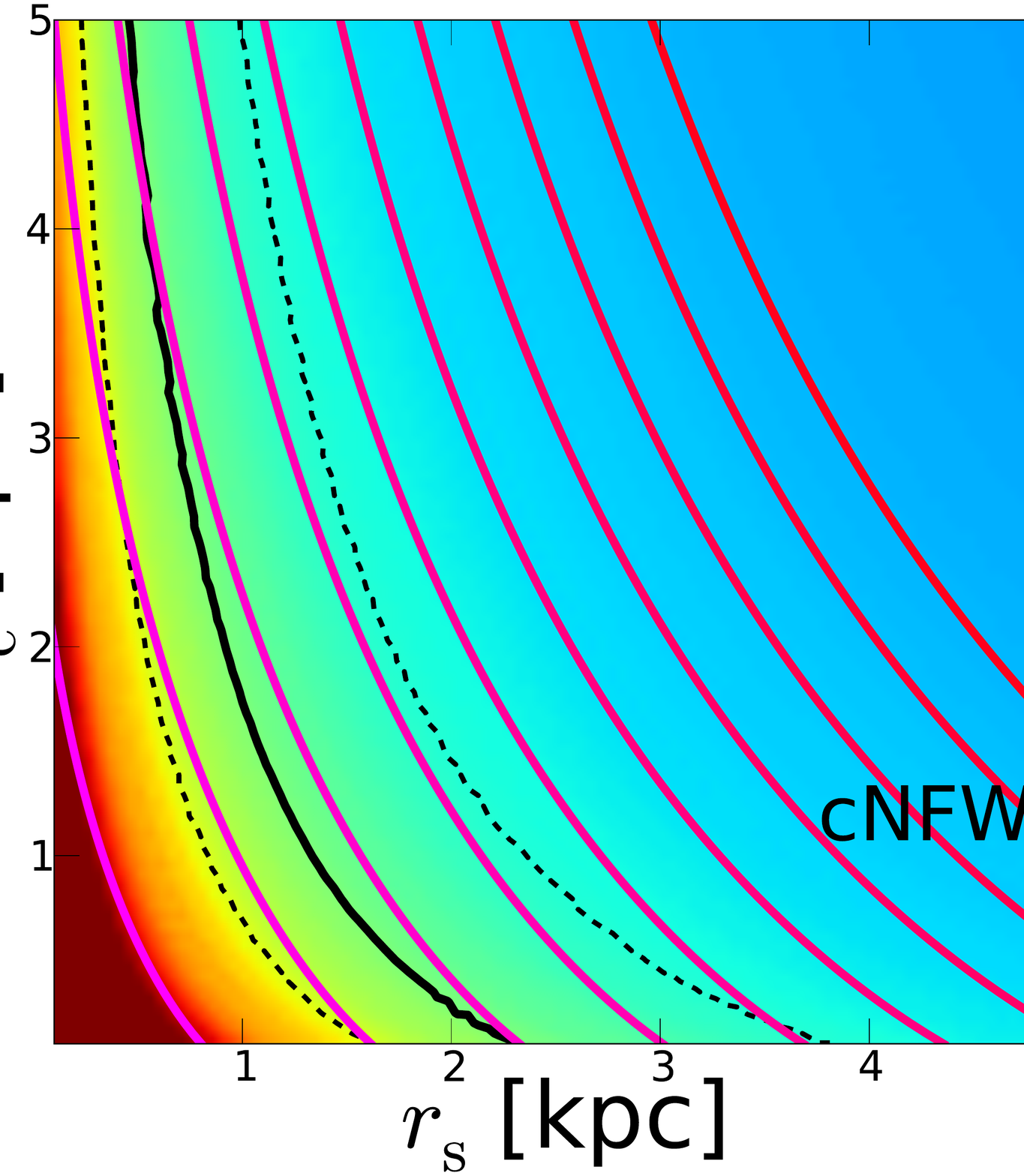}
\hspace{-0.1cm}\includegraphics[width=0.2\textwidth]{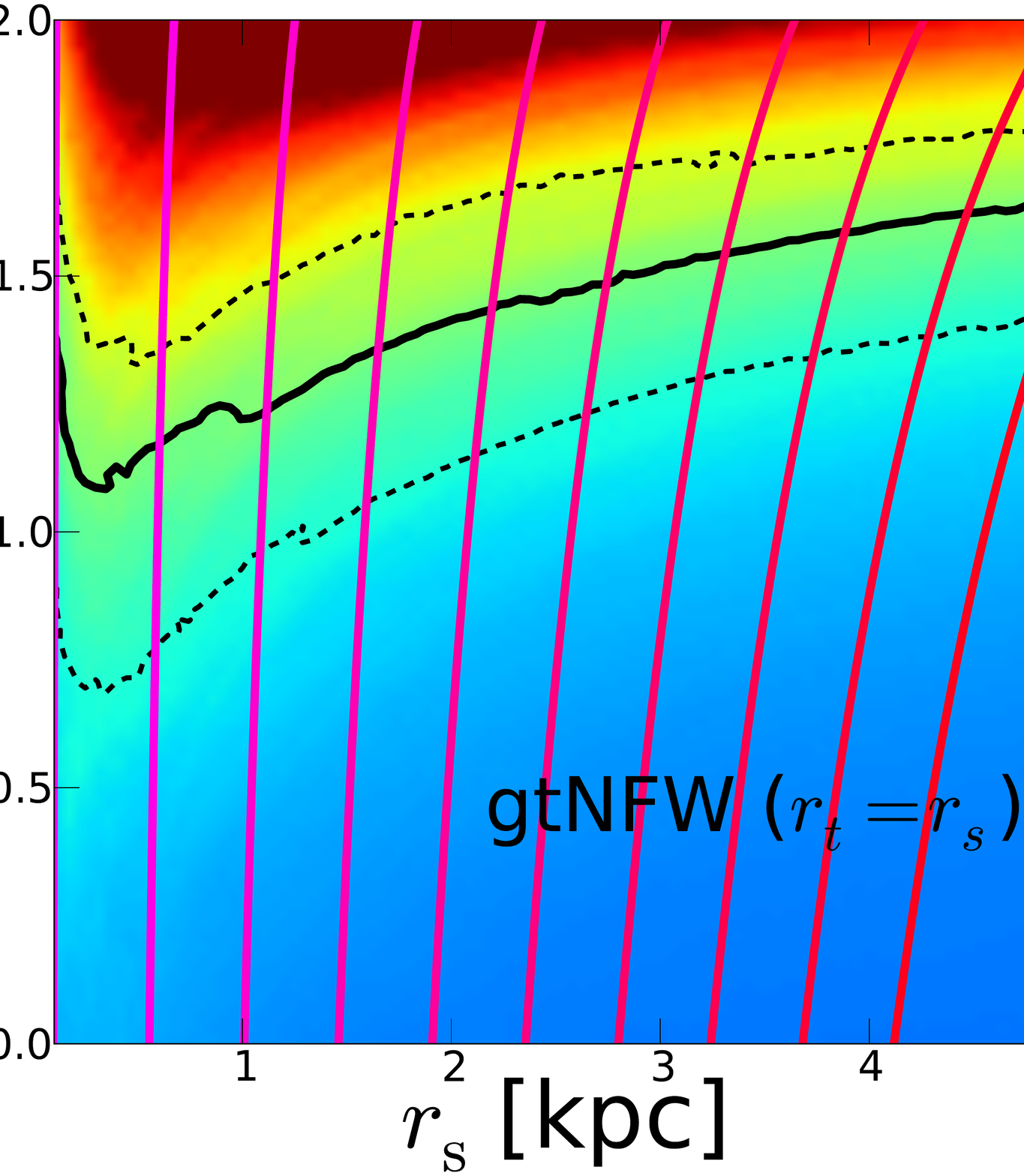}
\hspace{-0.1cm}\includegraphics[width=0.2\textwidth]{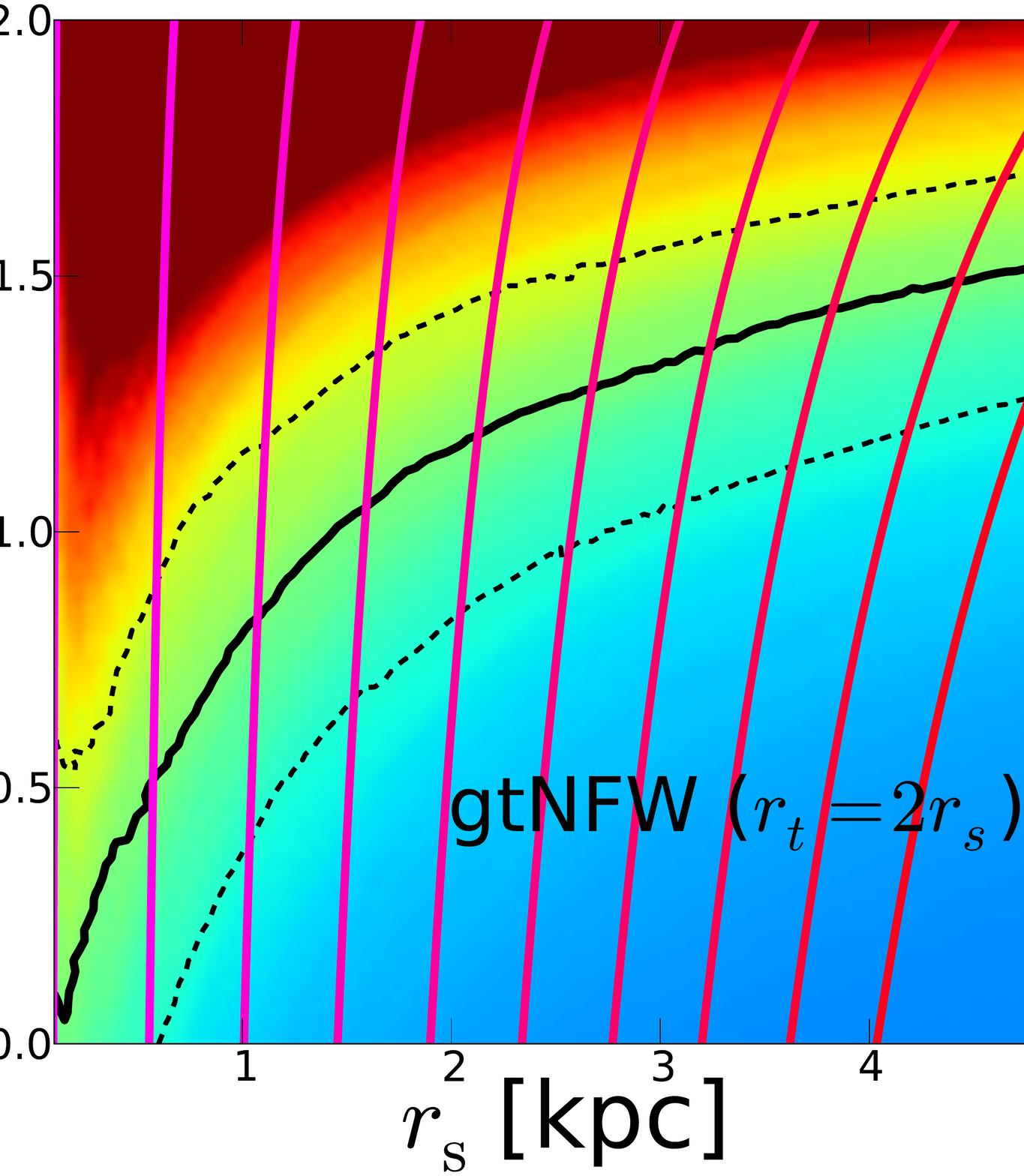}
\hspace{-0.1cm}\includegraphics[width=0.2\textwidth]{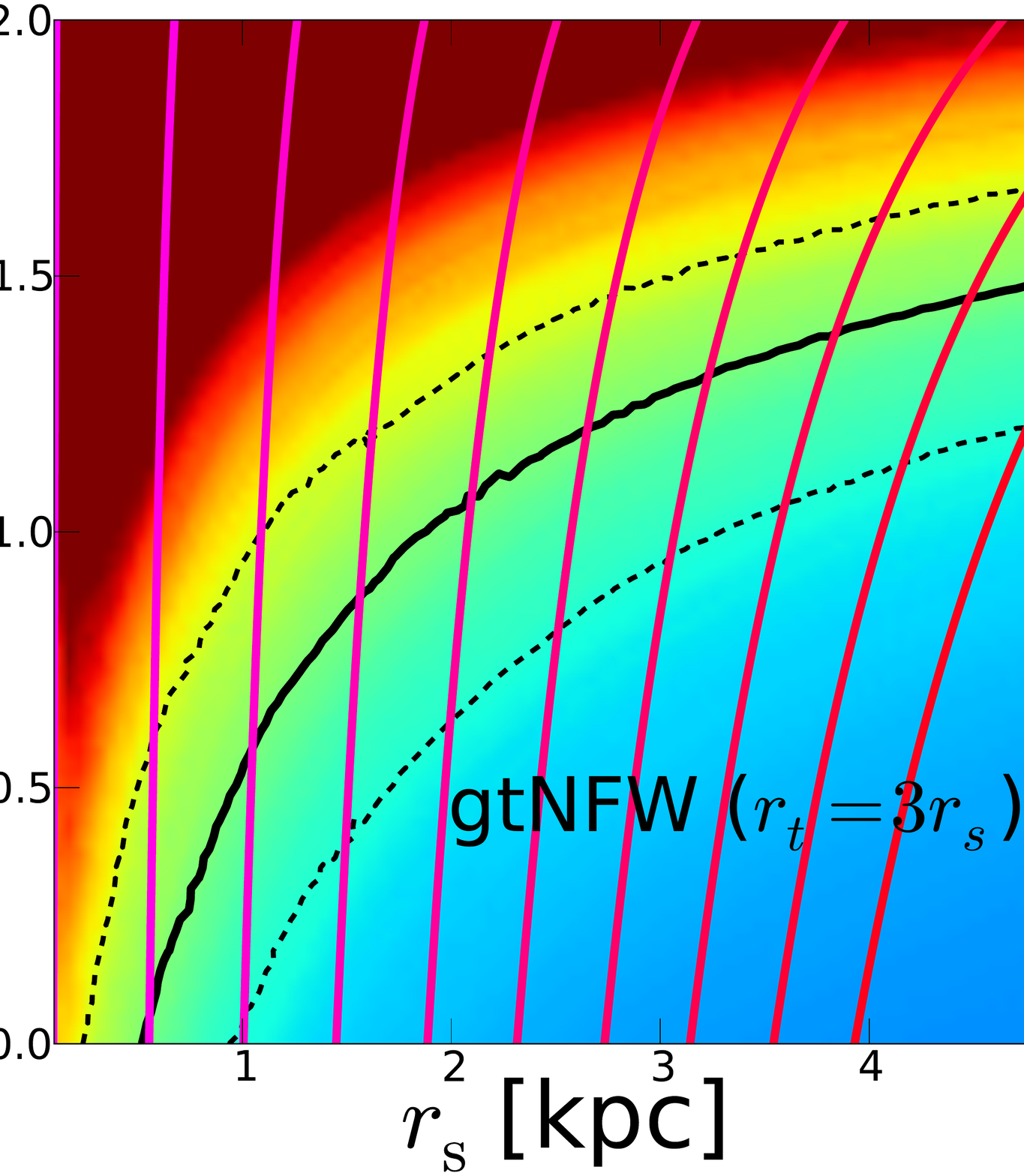}\\
\hspace{-0.1cm}\includegraphics[width=0.2\textwidth]{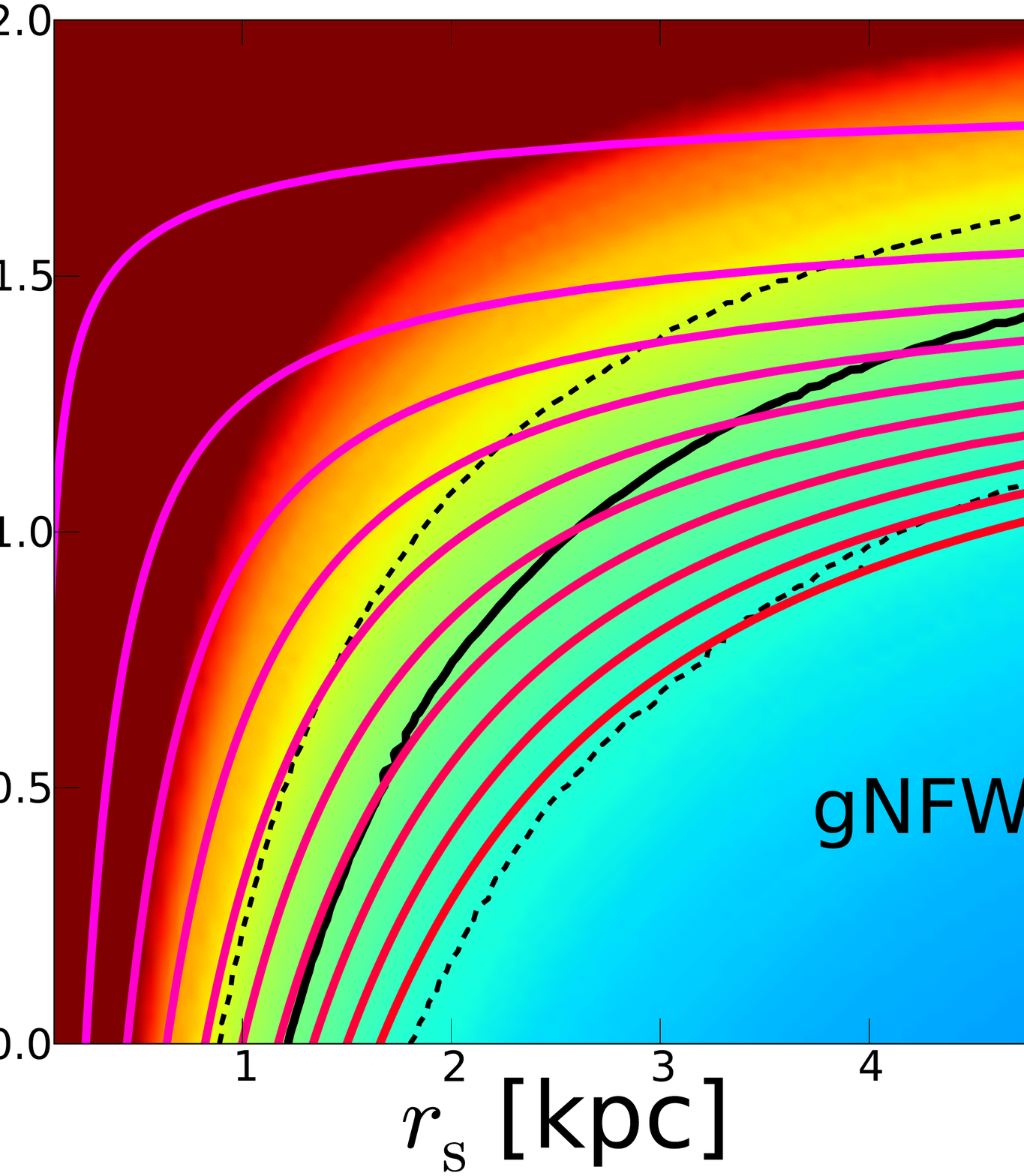}
\hspace{-0.1cm}\includegraphics[width=0.2\textwidth]{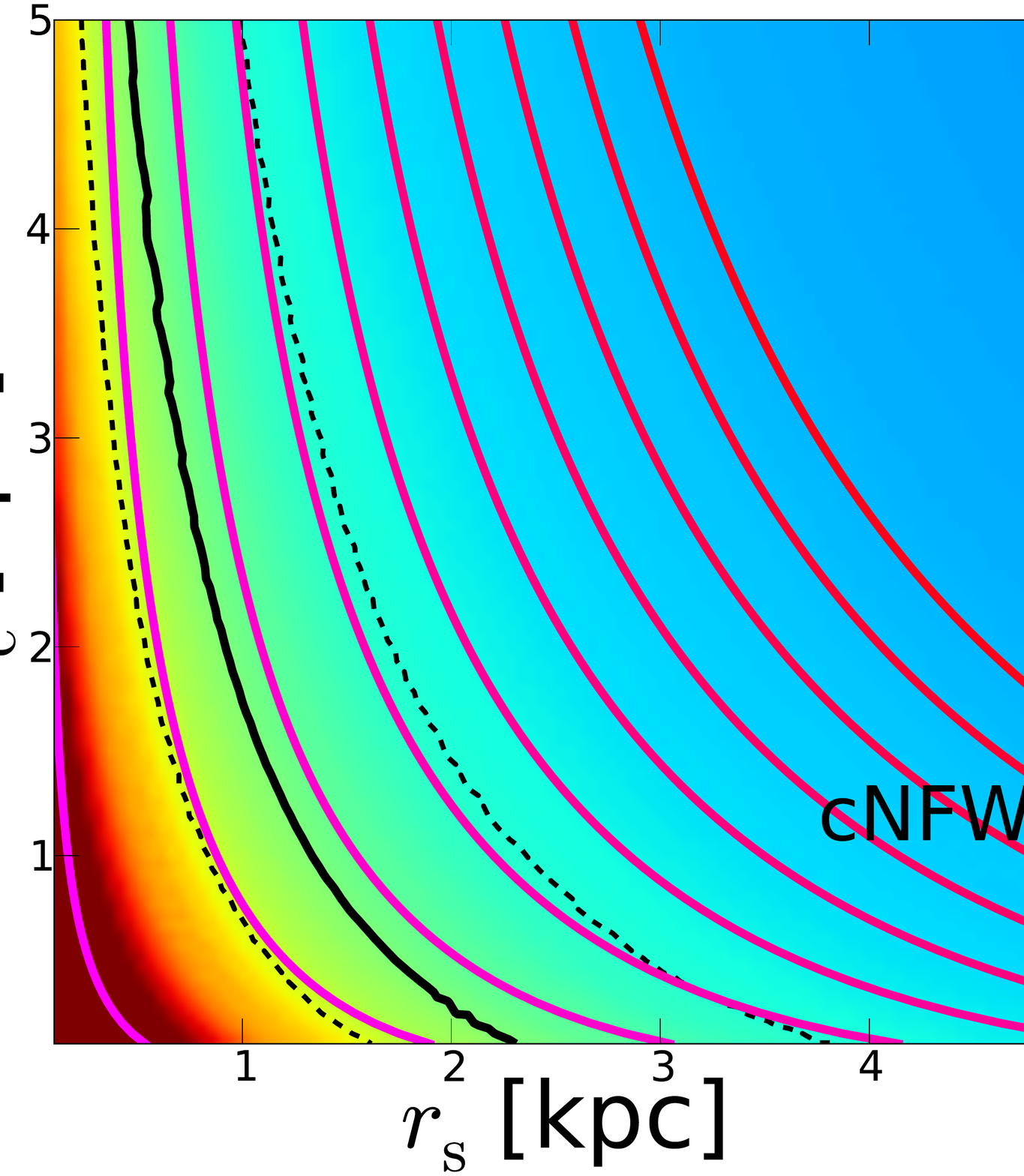}
\hspace{-0.1cm}\includegraphics[width=0.2\textwidth]{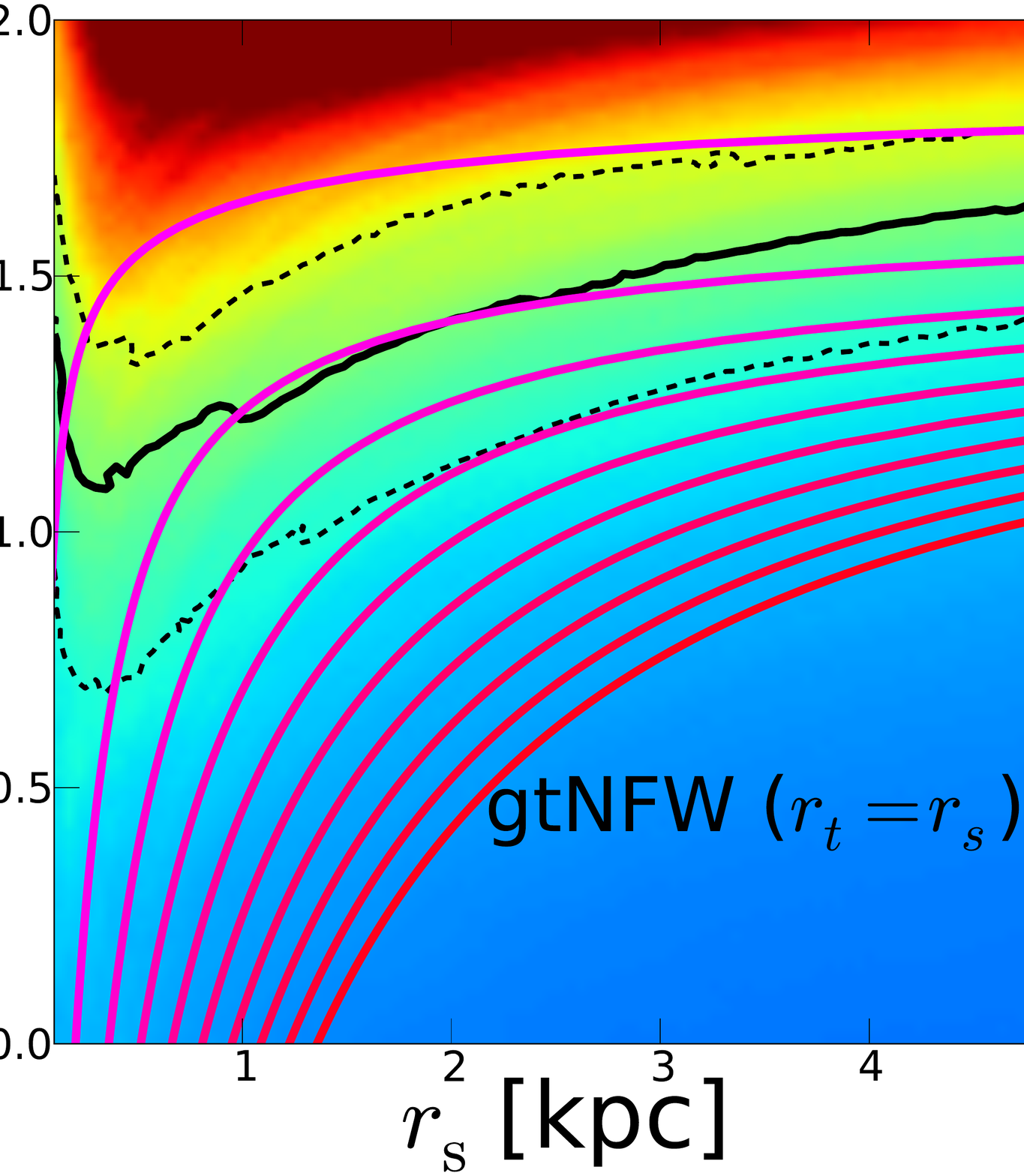}
\hspace{-0.1cm}\includegraphics[width=0.2\textwidth]{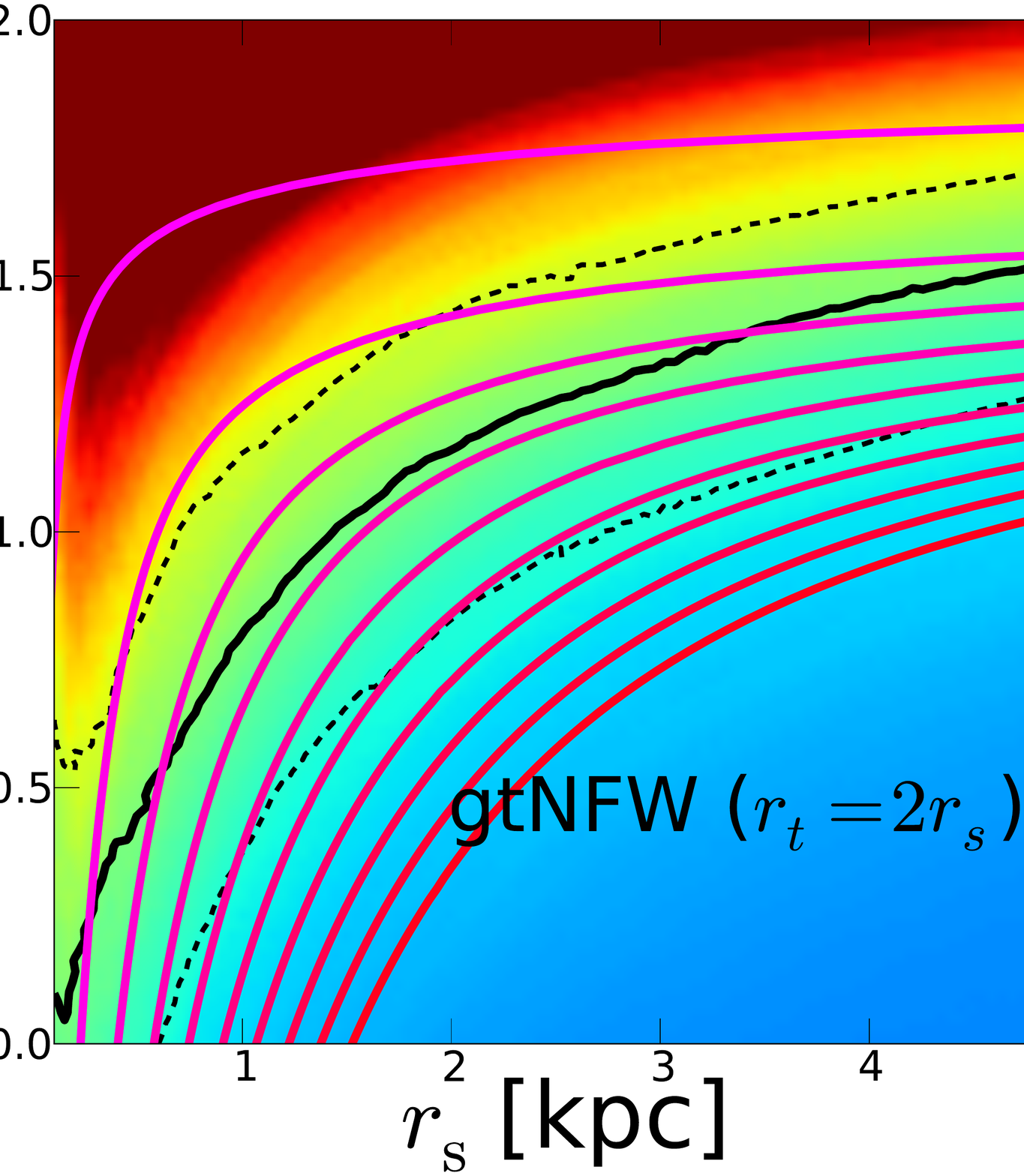}
\hspace{-0.1cm}\includegraphics[width=0.2\textwidth]{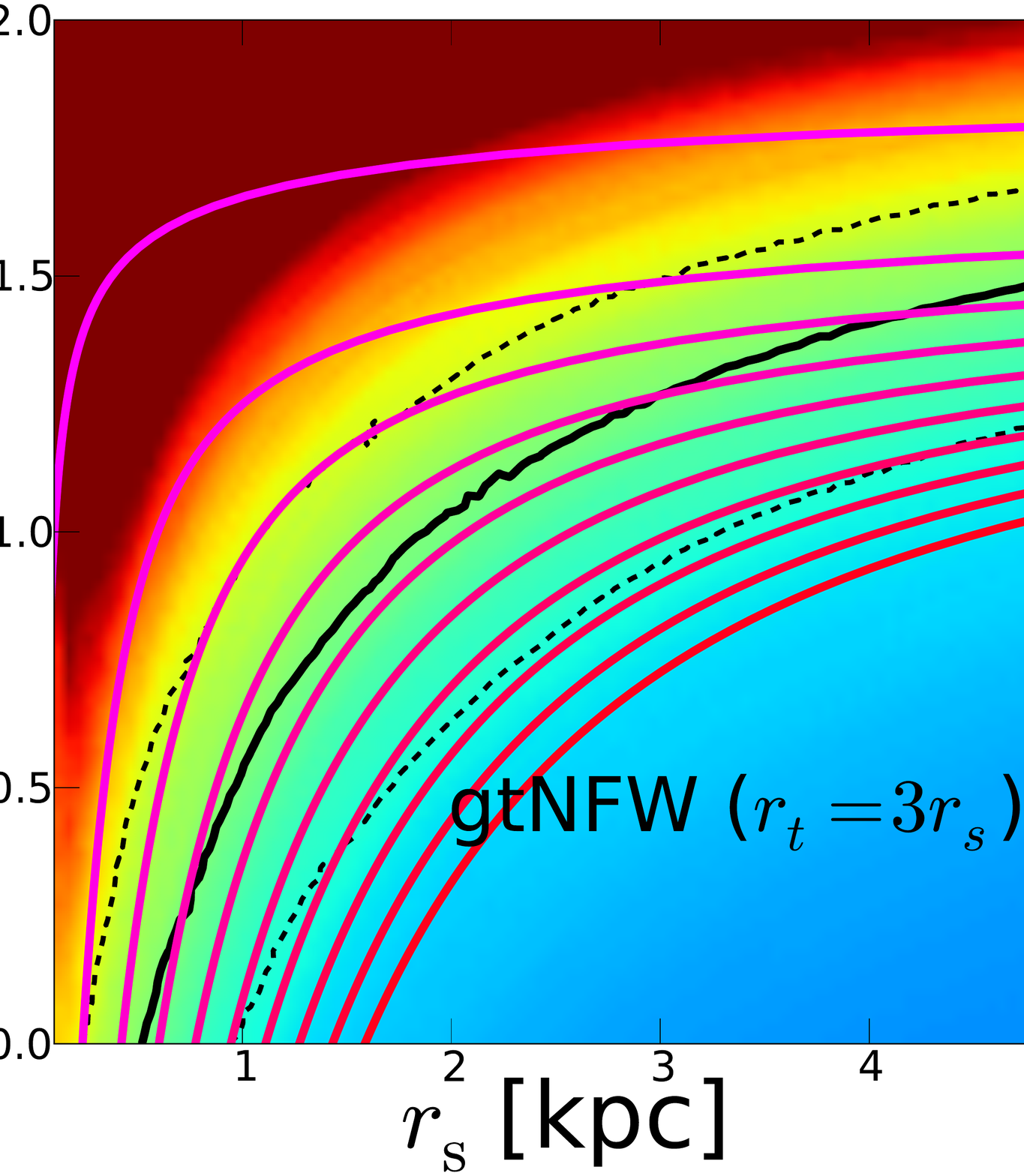}\\
\vspace{-0.1cm}\hspace{-0.1cm}\includegraphics[width=0.2\textwidth]{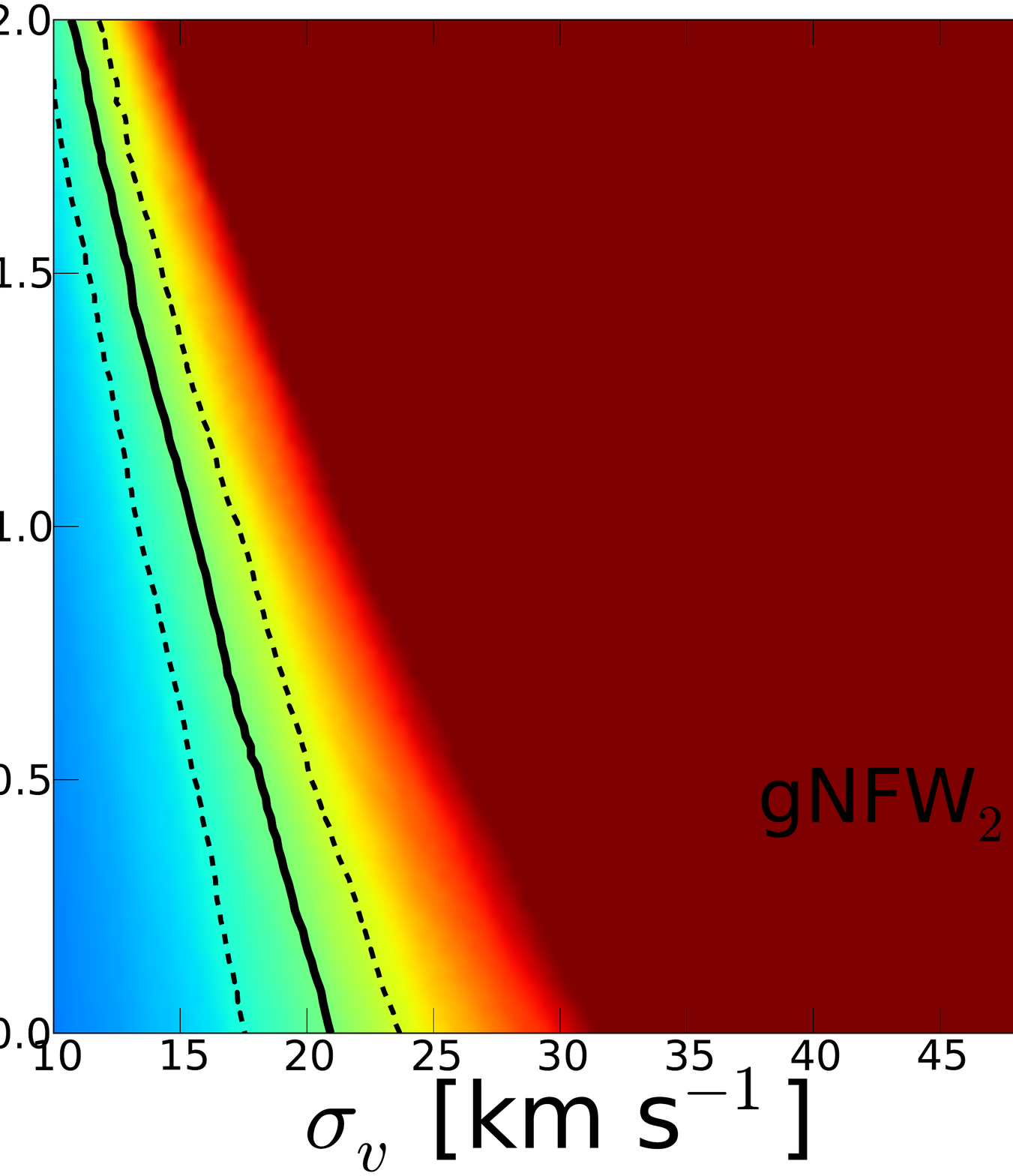}
\hspace{-0.1cm}\includegraphics[width=0.2\textwidth]{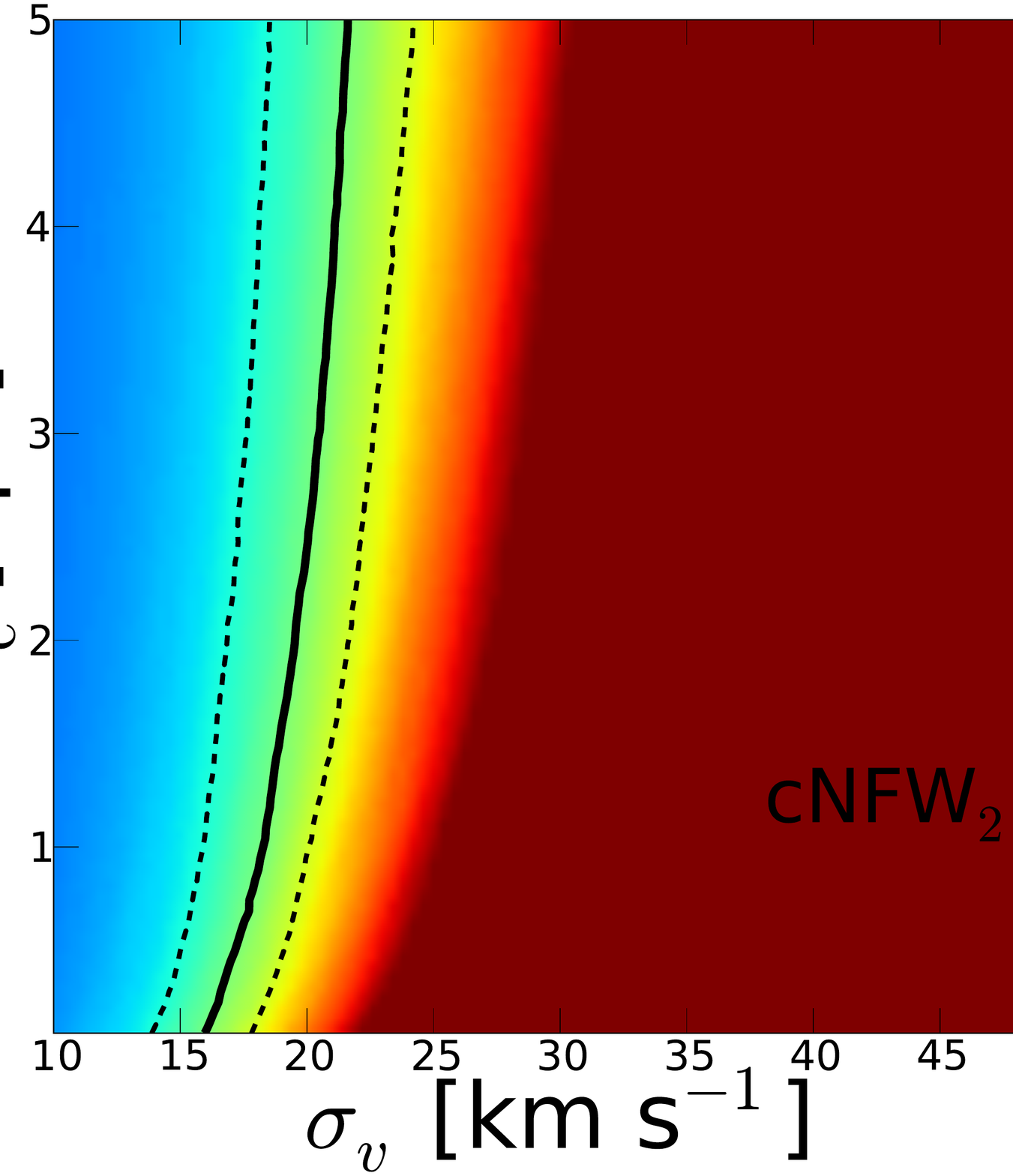}
\hspace{-0.1cm}\includegraphics[width=0.2\textwidth]{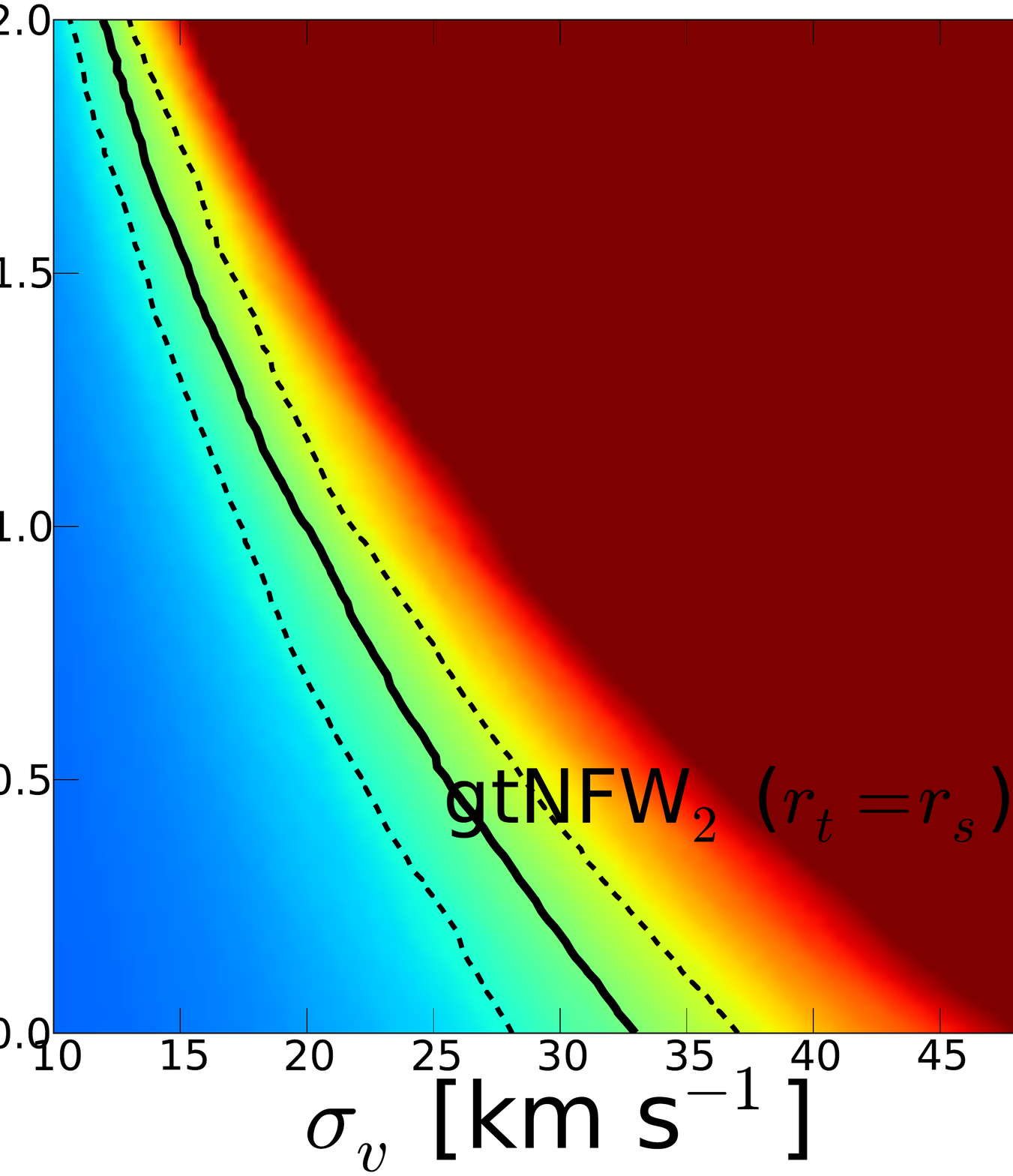}
\hspace{-0.1cm}\includegraphics[width=0.2\textwidth]{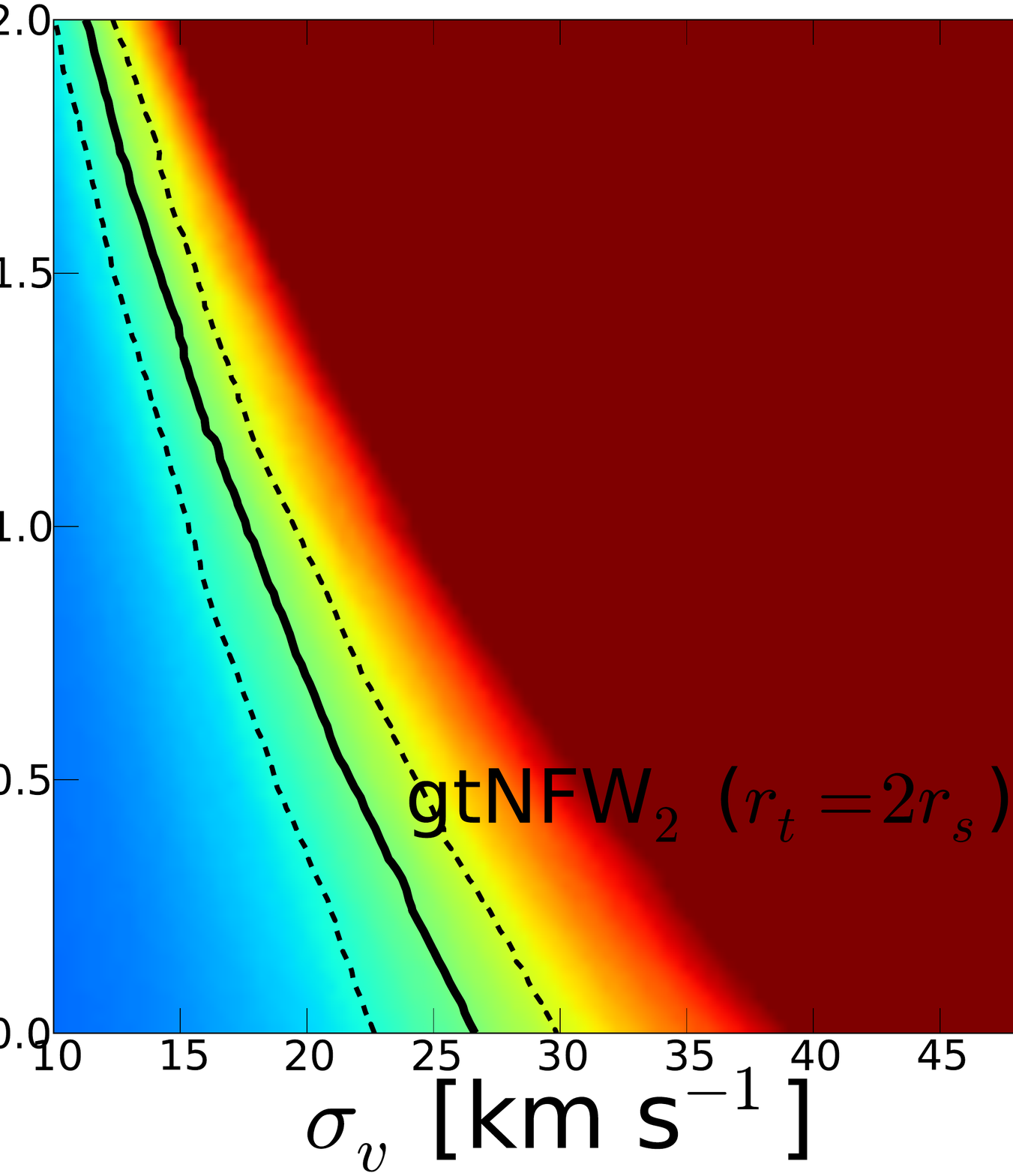}
\hspace{-0.1cm}\includegraphics[width=0.2\textwidth]{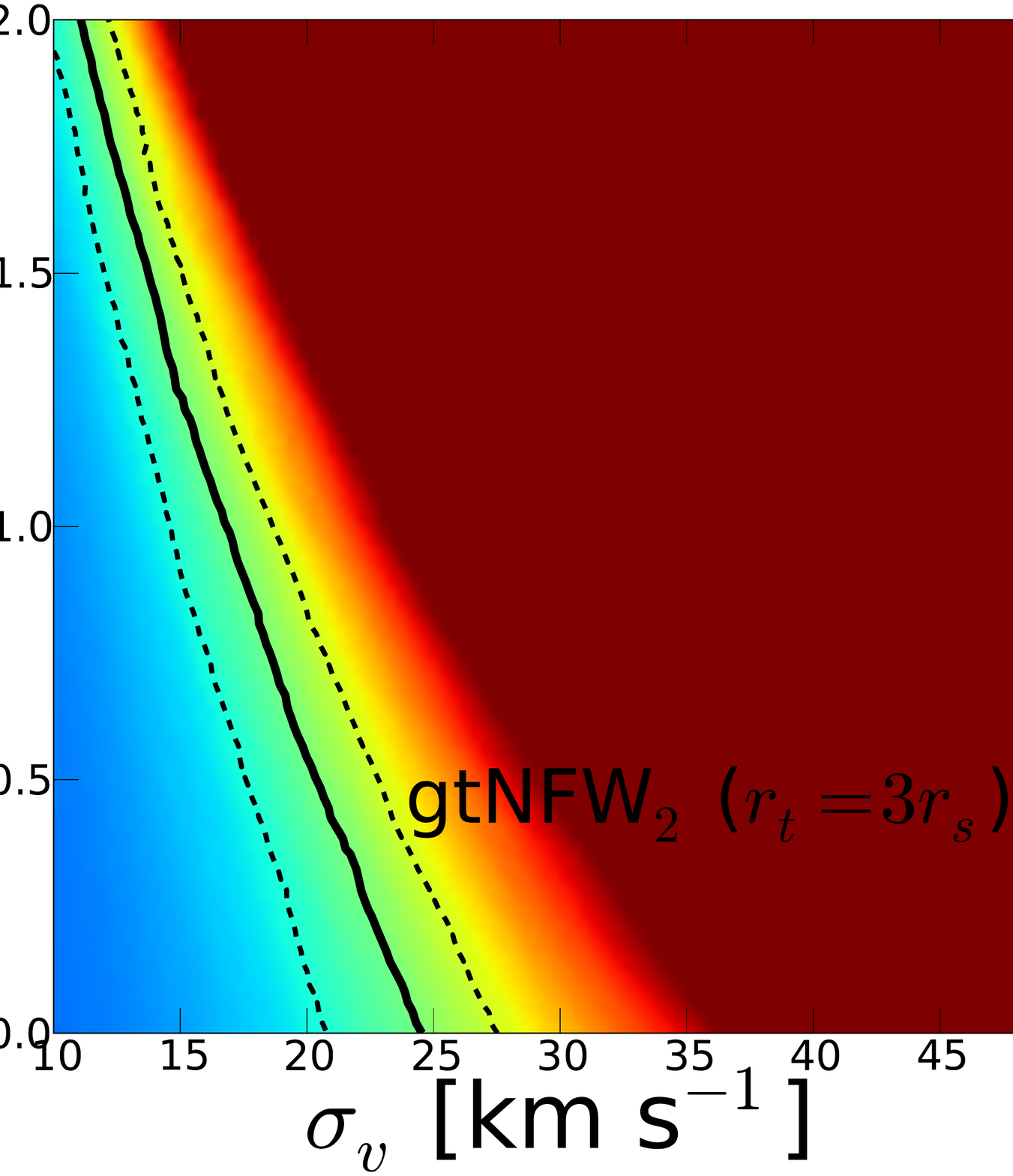}
\caption{The degeneracy among the mass model parameters on the lensing effect
{$\cal{D}$} of a substructure with different profiles (as indicated).  The
solid line indicates the region of the parameter space where the considered
substructure profile has the same lensing effect as a SIS (${\cal D}={\cal D}_{\rm SIS}$), while the dashed lines indicate the deviation from the solid line by
$\pm$~10 percent. The solid red lines are iso-$\chi^2_{rel}$ (top panels) and iso-$\chi^2$ (middle panels) lines and indicate the intrinsic degeneracy among the different mass profile parameters.
The colour scale is the quantity ${\cal D}$ normalized to ${\cal D}_{\rm SIS}$.}
\label{fig:profile_deg} 
\end{figure*}

\section{Discussion}
\label{sec:discussion}

\begin{table}
\begin{tabular}{lllll}
\hline
Profile         &$\sigma_v$ [$\mathrm{km\,s}^{-1}$]     &\!\!\!\!$r_{\rm{max}}$ [$\mathrm{kpc}$]   &$v_{\rm{max}}$ [$\mathrm{km\,s}^{-1}$]  &$r_t/r_s$\\
\hline
gNFW            &-                                   &\!\!\!\!~5.0\,$\pm$\,0.0                    &26.1\,$\pm$\,0.0                                                    &-\\

gNFW$_2$        &16.4\,$\pm$\,2.6                                  &-        &-                                                &-\\

gtNFW           &-                                   &\!\!\!\!0.9\,$\pm$\,0.4                   &31.4\,$\pm$\,5.1                                                   &1\\

gtNFW           &-                                  &\!\!\!\!1.9\,$\pm$\,0.8                   &28.9\,$\pm$\,4.8                                                  &2\\

gtNFW           &-                                   &\!\!\!\!2.6\,$\pm$\,0.9                  &27.1\,$\pm$\,5.1                                                   &3\\

gtNFW$_2$       &23.9\,$\pm$\,5.8                                  &-       				        &-                                                 &1\\

gtNFW$_2$       &19.8\,$\pm$\,4.3                                &-       				        &-                                                 &2\\

gtNFW$_2$       &18.8\,$\pm$\,3.7                                  &-       				        &-                                                 &3\\

cNFW            &-                                  &\!\!\!\!5.0\,$\pm$\,0.04                   &25.9\,$\pm$\,0.1                                                 &-\\

cNFW$_2$        &19.1\,$\pm$\,1.5                                &-                                &-                       &-\\
\hline
\end{tabular}
\caption{Derived kinematical properties for profiles with a lensing effect consistent with the observed SIS.}
\label{tbl:v_max}
\end{table}

The aim of this Paper is to investigate the gravitational lensing effect of different substructure profiles on the surface brightness distribution of extended gravitationally lensed images and determine which regions of the profile parameter space are in agreement or in disagreement with the observations.
To investigate these degeneracies, we show in Fig.~\ref{fig:profile_deg} the lensing effect of all of the considered profiles relative to the effect of a SIS profile, ${\cal D}/{\cal D}_{\rm SIS}$ (colour scale). The solid lines mark regions where the lensing signal of the profile under consideration is exactly the same as that expected for a SIS (${\cal D}/{\cal D}_{\rm SIS}=$~1). This line represents mass density profiles that have the same (low) likelihood to be fit by a smooth model than the reference SIS substructure. The dashed lines mark those regions within which the lensing signal deviates less than $\pm$~10~percent from the expected lensing signal of an SIS. What we see is that the lensing effect of the substructure as observed in JVAS\,B1938+666 is compatible with only a small combination of the different profile parameters, and that a large fraction of the parameter space can already be excluded.
Tests with different noise levels show that the signal-to-noise ratio of the data can broaden the $\pm 10$ percent region around the ${\cal D}/{\cal D}_{\rm SIS}=$~1 curve, while leaving the shape of the latter essentially unaffected.
This could break down in the case of significant covariance between the pixel noise; however we do not expect this to be the case for most ground-based adaptive optics observations.

We find that gNFW models with low (high) concentrations and shallow (steep) central slopes produce lensing effects that are too small (large) to provide a good fit to the data and that all of the considered values for the core radius are possible as long as the cNFW profile is concentrated enough. In a future paper, we will investigate whether this is an indication that substructure lensing is insensitive to the size of the core radius or whether this is an issue related to 
the smoothness of the source surface brightness distribution considered here. 
For highly truncated profiles (e.g. gtNFW with $r_t = r_s$), the radial extent of the substructure is significantly suppressed and the lensing signal becomes essentially only sensitive to the mass within the truncation radius. This implies that the lensing effect is in this case insensitive to the concentration as long as the slope is steep enough to provide enough mass in the central regions of the substructure. As the truncation radius increases, a larger range of slope is allowed, while the lensing
signal becomes slowly more sensitive to the scaling radius. 

Independently of the profile, we derive an upper limit on the mass density slope of $\gamma < 1.6$.

We now focus on the mass and kinematical properties of these profiles. Even though the lensing signal still shows an intrinsic degeneracy in the main model parameters, only profiles with a tight range of $r_{\rm{max}}$ and $v_{\rm{max}}$ combinations are consistent with the reference SIS. In particular, by considering all of the combinations of profile parameters that lead to ${\cal D}/{\cal D}_{\rm SIS}=$~1, we derive mean values of $r_{\rm{max}}=3.1$ $\mathrm{kpc}$ and $v_{\rm{max}}=28.5$ $\mathrm{km\,s}^{-1}$ (see Table \ref{tbl:v_max} for more details). We find, therefore, that the substructure properties of JVAS\,B1938+666 are consistent only with profiles that have kinematical properties within 1$\sigma$ from those of the bright dwarf spheroidal satellite galaxies of the Milky Way \citep{Boylan2011}. This is not an obvious result, since the host lens is a massive early-type galaxy at $z = 0.881$, and has therefore a mass which is much larger than the Milky Way. The bottom panels of Fig.~\ref{fig:profile_deg} show the lensing degeneracy among various profiles of different masses, but with the same concentration. From the ${\cal D}/{\cal D}_{\rm SIS}=$~1 regions on these plots, we find that that the substructure velocity dispersion is constrained to a relatively tight range between 13.8  $\mathrm{km\,s}^{-1}$ and  29.8 $\mathrm{km\,s}^{-1}$ ( see Table \ref{tbl:v_max} for more details).  
From all the profiles with ${\cal D}/{\cal D}_{\rm SIS}=$~1, we then derive an average mass within 300 pc of $M_{300}=$~8.3~$\times$~10$^7~M_\odot$ with an rms of $1.4\times~10^8~M_\odot$. This is consistent with the mass of the substructure found in JVAS\,B1938+666, as measured by \citet{Vegetti12} under the assumption of a pseudo-Jaffe profile, $M_{300}=$~1.13\,$\pm$\,0.06~$\times$~10$^7~M_\odot$. Within each of the different profiles, we find smaller scatters and larger errors (relative to a SIS) for the substructure mass, but these are always consistent with the results of \citet{Vegetti12}. Finally, we derive an average projected mass within the Einstein of $M_{\rm E}=3.2\times 10^6M_\odot$. This is also comparable with the mass within the Einstein radius of the reference SIS $M_{\rm SIS}= 4\times 10^6M_\odot$. This indicates that substructure lensing provides a reliable measure of the main substructure properties independently on the assumed profile and that the gravitational imaging technique provides a precise measure of the lensing mass. We can therefore conclude that the substructure mass function derived with the gravitational imaging technique is not biased by our assumption on the form of the substructure density profile at least relative to the detections.\\
Different dark-matter models make different predictions for the density profile of galaxy haloes and sub-haloes. In order to compare these theoretical expectations with the gravitational lensing observations, it is important to understand how the corresponding degeneracies behave relative to each other. The top and middle panels of Fig.~\ref{fig:profile_deg} show the lensing degeneracy along with the intrinsic profile degeneracy that is obtained by minimizing the $\chi^2_{rel}$ and the $\chi^2$, respectively. We find that those predictions that are based on fits dominated by the density profile at large radii could significantly over- or under-estimate the lensing effect and, therefore, the predicted 
number of substructures that are observable with gravitational lensing. This is particularly true for truncated profiles, where the two forms of degeneracies become almost orthogonal to each other.

\section{Conclusions}
\label{sec:conclusions}
In this Paper, we have investigated the gravitational lensing effect of different substructure models on highly magnified Einstein rings. Our main results can be summarized as follows: (i) the gravitational imaging technique can be used to exclude large regions of the considered parameter space and therefore models that predict a large number of satellites in those regions; (ii) only profiles with the right level of central concentration provide a good fit to the data; (iii) given the signal-to-noise ratio and angular resolution of the data, and the source surface brightness distribution considered here, the gravitational lensing effect of mass substructure is essentially insensitive to the size of cores; (iv) even at the substructure level gravitational lensing provides a precise measure of the lensing mass; (v) independent of the assumed mass profile, substructure observations in the gravitational lens galaxy B1938+666 are consistent with values of $r_{\rm{max}}$ and $v_{\rm{max}}$ that are within 1$\sigma$ of the values derived for the luminous dwarf satellites of the Milky Way and a mass density slope $\gamma < 1.6$; (vi) a theoretical prediction based on fits that are dominated by the density profile at larger radii may significantly over or under-estimate the number of substructure detectable with lensing.

In light of these results we can conclude that substructure lensing provides an important tool to 
explore the density profiles of galaxy satellites beyond the local Universe.
We stress once more that the results presented in this Paper are based on specific assumptions on the substructure mass and location on the arc and
on the substructure sensitivity. We have also made specific assumptions on the observational quality of the lensed images to those made with ground based adaptive optics on a 10-m class telescope (see Lagattuta et al. 2012 and Vegetti et al. 2012).
We refer to a future paper for a wider analysis that takes into account more general physical and observational scenarios. Thanks to the increased angular resolution, we expect to be able to put tighter constraints on the number of allowed substructure profiles in the near term using very long baseline interferometry at radio wavelengths of extended arcs, and in the more distant future with the E-ELT.

\section*{Acknowledgements}
We thank Phil Marshall, Jesus Zavala, Simon White and the anonymous referee for useful discussions. 
\bibliography{ms}

\end{document}